\documentclass[fp]{jpsj3}
\usepackage{graphicx,psfrag}
\usepackage{graphicx} 
\usepackage{bm}
\usepackage{booktabs}
\usepackage{amsmath,amsthm,amssymb}
\usepackage{fancybox} 
\usepackage{ascmac} 
\usepackage{framed}

\usepackage{color}
\begin{document}

\title{
Coupling Theory of Emergent Spin Electromagnetic Field and Electromagnetic Field
}

\author{Hideo Kawaguchi$^1$\thanks{E-mail:kawaguchi-hideo@ed.tmu.ac.jp} and Gen Tatara$^2$}
\inst{$^1$ Graduate School of Science and Engineering, Tokyo Metropolitan University, Hachioji, Tokyo 192-0397, Japan\\
$^2$ RIKEN Center for Emergent Matter Science (CEMS), Wako, Saitama 351-0198, Japan}

\date{\today}
\abst{ \ \ \ 
In ferromagnetic metals, an  effective electromagnetic field that couples to conduction electron spins is induced by the $sd$ exchange interaction. 
We investigate how this effective field, namely, the spin electromagnetic field, interacts with the ordinary electromagnetic field by deriving an effective Hamiltonian based on the path integral formalism.
It turns out that the dominant coupling term is the product of the electric field and spin gauge field. This term describes the spin-transfer effect, as was pointed out previously.
The electric field couples also to the spin electric field, but 
this contribution is smaller than the spin-transfer contribution in the low frequency regime.
The magnetic field couples to the spin magnetic field, and this interaction suggests an intriguing intrinsic mechanism of frustration in very weak metallic ferromagnets under a uniform magnetic field.
We also propose a voltage generation mechanism due to a nonlinear effect of non-monochromatic spin-wave excitations.
}

\newcommand{\Asv}{{\bm A}_{\rm s}}
\newcommand{\Bsv}{{\bm B}_{\rm s}}
\newcommand{\Esv}{{\bm E}_{\rm s}}
\newcommand{\ev}{\bm{e}}
\newcommand{\nv}{\bm{n}}
\newcommand{\qv}{\bm{q}}
\newcommand{\kv}{\bm{k}}
\newcommand{\rv}{{\bm r}}

\newcommand{\Sv}{{\bm S}}
\maketitle

\section{Introduction}
Spin-transfer torque induced by an applied electric current in ferromagnetic metals is a crucially important  effect in spintronics.
The idea was first proposed theoretically by Berger \cite{Berger86} in the case of a domain wall motion and by Slonczewski \cite{Slonczewski96} and Berger \cite{Berger96} in the case of the uniform magnetization of thin films.
The spin-transfer effect arises from the transfer of spin angular momentum from conduction electrons to localized spins which induce the magnetization.
The effect is caused by the $sd$ exchange interaction, and the angular momentum transfer occurs owing to the angular momentum conservation \cite{Slonczewski96}.
The interaction Hamiltonian describing the spin-transfer effect is 
\begin{align}
  \mathcal{H}_{\rm st}&= \int {\rm{d}}^3r \frac{\hbar P}{2e}(1-\cos\theta) (\bm{j}\cdot\nabla)\phi,
\end{align}
where $\theta$ and $\phi$ are the polar coordinates representing the localized spin direction, $\bm{j}$ denotes the applied electric current density, $P$ is the spin polarization of the conduction electron, and $e$ is the electron charge.
The interaction is represented as a gauge coupling to a spin gauge field $\bm{A}_{\rm s}^z$ 
\cite{Bazaliy98,TKS_PR08}, 
$\mathcal{H}_{\rm st}= \int {\rm{d}}^3r (\bm{j}_{\rm s}\cdot\bm{A}_{\rm s}^z)$, where $\bm{j}_{\rm s}\equiv P\bm{j}$ and $\bm{A}_{\rm s}^z=\frac{\hbar}{2e}(1-\cos\theta) \nabla\phi$.
The interaction is thus expressed as 
$\mathcal{H}_{\rm st}= \int {\rm{d}}^3r P\sigma_{\rm B}(\bm{E}\cdot\bm{A}_{\rm s}^z)$, where $\sigma_{\rm B}$ is the Boltzmann conductivity.
This expression clearly shows that the spin-transfer effect is due to a coupling of two gauge fields, the conventional electromagnetic field of the electric charge, and the gauge field acting on the electron spin.
The aim of this work is to study the coupling between the two gauge fields by calculating an effective Hamiltonian. 
We shall show that the effective Hamiltonian in the case of a slowly varying magnetization is made up of three contributions, one representing the spin-transfer torque and the others describing the couplings between the electric and magnetic fields.

In the case of charge electromagnetism coupled to relativistic charged particles, the effective Lagrangian induced by the particles is always written in a relativistically invariant form as $\sum_{\mu\nu}F_{\mu\nu}F^{\mu\nu}$, where $F_{\mu\nu}$ is the field strength and $\mu$ and $\nu$ are indices representing $x,y,z$, and $t$. The only terms allowed in the relativistic case are thus proportional to either $|\bm{E}|^2$ or $|\bm{B}|^2$. 
In ferromagnetic metals, conduction electrons interact with two gauge fields, $\bm{A}$ acting on the charge and $\bm{A}_{\rm s}$ acting on the spin, and the total electric and magnetic fields become $\bm {E}+\bm{E}_{\rm s}$ and $\bm {B}+\bm{B}_{\rm s}$, where  $\bm{E}_{\rm s}$ and $\bm{B}_{\rm s}$ are the effective spin electric and spin magnetic fields, respectively. If the system is relativistic, we would thus expect to have interactions in the form of $\bm {E}\cdot\bm{E}_{\rm s}$ and $\bm {B}\cdot\bm{B}_{\rm s}$ arising from $(\bm {E}+\bm{E}_{\rm s})^2$ and $(\bm {B}+\bm{B}_{\rm s})^2$. In reality, there are other contributions in ferromagnetic metals since the electrons are not relativistic and they have a finite lifetime of elastic scattering. 
We shall demonstrate that a coupling term proportional to $\bm{E}\cdot \bm{A}_{\rm s}^z$ arises as the dominant contribution. This term derived first in Ref. 6 represents the spin-transfer effect, as was discussed there. We also investigate other coupling terms, $\bm {E}\cdot\bm{E}_{\rm s}$ and $\bm {B}\cdot\bm{B}_{\rm s}$.

\subsection{Spin electromagnetic field}
An effective electromagnetic field arises from the $sd$ exchange interaction 
described by 
\begin{align}
  \mathcal{H}_{sd}=-\Delta_{sd}\int {\rm{d}}^3r \bm{n}\cdot \bm{s}_{\rm e}, 
\end{align}
where $\Delta_{sd}$ is the exchange energy, $\bm{n}$ is a unit vector representing the direction of the localized spin, and  $\bm{s}_{\rm e}$ is the direction of the conduction electron spin.
When this exchange interaction is strong, the conduction electron spin is aligned parallel to the localized spin direction, and this effect results in a quantum mechanical phase attached to the electron spin when the electron moves (see Ref. 7 for details of derivation). 
The spin part of the electron wave function with the expectation value along $\bm{n}$ is 
$|\bm{n}\rangle =\cos\frac{\theta}{2}|\uparrow\rangle+\sin\frac{\theta}{2}e^{i\phi}|\downarrow\rangle$,
where $\theta$ and $\phi$ are the polar coordinates of $\bm{n}$ and 
$|\uparrow\rangle$ and $|\downarrow\rangle$ denote the spin states \cite{Sakurai93}.
When the electron hops over a small distance ${\rm{d}}\rv$ to a nearby site 
where the localized spin is along $\bm{n}'$, the overlap of the wave functions is calculated as 
$\langle \nv'|\nv\rangle\simeq e^{\frac{i}{\hbar}e\Asv^{z}\cdot {\rm{d}}\rv}$, where 
\begin{equation}
 \Asv^{z}=\frac{\hbar}{2e}(1-\cos\theta)\nabla\phi, \label{Asdef}
\end{equation}
and the factor of $\frac{1}{2}$ is due to the magnitude of the electron spin.
The field $\Asv^{z}$ is an effective vector potential or an effective gauge field.
When the electron's path is finite, the phase becomes 
$
\varphi=\frac{e}{\hbar} \int_C {\rm{d}}\rv\cdot \bm{A}_{\rm s}^z
$. 
The existence of the phase means that there is an effective magnetic field $\Bsv$, as seen by rewriting the integral over a closed path using the Stokes theorem as
$
\varphi=\frac{e}{\hbar} \int_S {\rm{d}}\Sv\cdot\Bsv
$,
 where
$\Bsv\equiv \nabla\times\bm{A}_{\rm s}^z$.
The time derivative of the phase is equivalent to a voltage, and thus, we have an effective electric field defined by
$
\dot{\varphi}=-\frac{e}{\hbar} \int_C{\rm{d}}\rv\cdot\Esv
$, 
 where $\Esv\equiv-\dot{\bm{A}}_{\rm s}^z$.
These two fields satisfy Faraday's law,
$
\nabla\times\Esv+\dot{\Bsv}=0
$.
We therefore have effective electromagnetic fields that couple to the conduction electron spin as a result of the $sd$ exchange interaction.
We call the field a spin electromagnetic field \cite{Tatara_smf13}.
Using the explicit form of the effective gauge field, Eq. (\ref{Asdef}), we see that the emergent spin electromagnetic fields are
\begin{eqnarray}
{\Esv}_{,i}&=&-\frac{\hbar}{2e} \nv \cdot (\dot{\nv} \times \nabla_i \nv), \nonumber\\
{\Bsv}_{,i}&=& \frac{\hbar}{4e}\sum_{jk}\epsilon_{ijk} \nv \cdot (\nabla_j \nv \times \nabla_k \nv) .\label{EBtopological}
\end{eqnarray}
The magnetic component ${\Bsv}$ is the spin Berry's curvature \cite{Xiao10} or scalar chirality.
The electric component ${\Esv}$, called the spin motive force, is a chirality in the space-time, which arises when the localized spin structure $\bm{n}$ is time-dependent.

The expression Eq. (\ref{EBtopological}) was derived by Volovik in 1987 \cite{Volovik87}.
Originally, the emergence of the effective electric field $\Esv$ from moving magnetic structures was found in 1986 by Berger, where a voltage generated by canting a moving domain wall was calculated \cite{Berger86}.
Stern discussed the motive force in the context of the spin Berry's phase and the Aharonov-Bohm effect in a ring, and showed similarity to Faraday's law \cite{Stern92}.
The spin motive force was rederived in Ref. 13 in the case of the domain wall motion, and discussed in the context of topological pumping in Ref. 14.
Those works consider only the adiabatic limit, i.e., in the case of a strong $sd$ exchange interaction and in the absence of spin-dependent scattering.
The idea of the spin motive force has recently been extended to include the spin-orbit interaction  \cite{Duine08,Lucassen11,Shibata09,Shibata11,Kim12,Tatara_smf13,Nakabayashi14}, and it was shown that the spin-orbit interaction modifies the spin electric field.
It was also shown that the spin electromagnetic field arises even in the limit of a weak $sd$ interaction \cite{Takeuchi12,Tatara12}. 
The case of the Rashba spin-orbit interaction has been studied in detail recently. It was shown that the spin electric field in this case emerges even from a uniform precession of magnetization \cite{Tatara_smf13,Nakabayashi14}. 
This fact suggests that the Rashba interaction at interfaces would be useful in controlling the spin-charge conversion.
The Rashba-induced spin electric field induces a voltage in the same direction as in the inverse spin Hall and inverse Edelstein effects \cite{Saitoh06,Sanchez13} driven by the spin pumping effect \cite{Tserkovnyak02}. 
It was also pointed out that the spin electromagnetic fields in the presence of spin relaxation satisfy Maxwell's equations with spin magnetic monopoles that are driven dynamically \cite{Takeuchi12}.
The coupling between the spin magnetic field and the helicity of light was theoretically studied in the context of the topological inverse Faraday effect, which is a nonlinear effect with respect to the incident electric field \cite{Taguchi12}. 

Experimentally, the spin magnetic field (the spin Berry's curvature) has been observed using the anomalous Hall effect in frustrated ferromagnets \cite{Lee09, Nagaosa10}. 
The spin electric field has been measured in the motion of various ferromagnetic structures such as domain walls \cite{Yang09}, magnetic vortices \cite{Tanabe12}, and skyrmions \cite{Schulz12}.

\section{Derivation of Effective Hamiltonian}

The effective Hamiltonian is calculated in the imaginary-time (denoted by $\tau$) path integral formalism \cite{Sakita85}. In this section, we set $\hbar = 1$.
The system we consider is a ferromagnetic metal, where conduction electrons, represented by two-component annihilation and creation fields, ${c} (\bm{r},\tau)$ and $\bar{c} (\bm{r},\tau)$, interact with localized spins, described by the vector field $\bm{n}(\bm{r},\tau)$, via the $sd$ exchange interaction.
The Hamiltonian thus reads
\begin{align}
\mathcal{H} &= \mathcal{H}_{\text{0}}  +\mathcal{H}_{sd} + \mathcal{H}_{\text{em}} , \\
\mathcal{H}_{\text{0}}  &=  \int {\rm{d}}^{3} r \left(\frac{1} {2m} |\bm{\nabla}{c} (\bm{r},\tau)|^2 - \mu \bar{c} (\bm{r},\tau) c (\bm{r},\tau)\right) ,\nonumber\\ 
\mathcal{H}_{sd} &= -\Delta_{sd} \int {\rm{d}}^{3} r   \bm{n} (\bm{r},\tau) \cdot \left(\bar{c} (\bm{r},\tau) \bm{\sigma} c (\bm{r},\tau)\right), 
\end{align}
where $\mu$ is the chemical potential, $m$ is the electron mass, and $\bm{\sigma}$ is the vector of Pauli matrices.
The term $\mathcal{H}_{\text{em}}$ represents the interaction between the conduction electron and the applied electromagnetic field, described by a vector potential $\bm{A}$, which reads 
\begin{align}
\mathcal{H}_{\text{em}} &= - \int {\rm{d}}^{3} r \bm{A}(\bm{r},\tau) \cdot \left(\frac{ie}{2m} \bar{c} (\bm{r},\tau) \overleftrightarrow{\nabla} c (\bm{r},\tau) - \frac{e^2}{2m}  \bm{A}(\bm{r},\tau) \bar{c} (\bm{r},\tau) c (\bm{r},\tau)\right),
\end{align}
where $\bar{c} \overleftrightarrow{\nabla} c $ $\equiv$ $\bar{c}\left(\bm{\nabla} c \right)-\left(\bm{\nabla} \bar{c} \right)c $ and $-e$ is the electron charge ($e>0$). The system we consider is a film thinner than the penetration depth of the electromagnetic field. 
The Lagrangian of the system is 
\begin{align}
\mathcal{L}&= \int {\rm{d}}^{3} r \bar{c} (\bm{r},\tau) \partial_\tau c(\bm{r},\tau)+\mathcal{H},
\end{align}
and the effective Hamiltonian describing the localized spin and the gauge field is obtained by carrying out a path integral over the conduction electrons as 
$\mathcal{H}_{\rm eff}(\theta,\phi,\bm{A})\equiv -\ln Z$, where 
\begin{align}
Z & = \int D\bar{c}(\bm{r},\tau)Dc(\bm{r},\tau)e^{-\int_0^\beta d\tau \mathcal{L}},
\end{align}
is the partition function and $D$ denotes the path integral.

We are interested in the case where the $sd$ exchange interaction is large and thus the conduction electron spin is aligned parallel to the localized spin direction $\bm{n}$, i.e., the adiabatic limit.
To describe this limit, the use of the spin gauge field, which characterizes the deviation from the adiabatic limit, is convenient \cite{TKS_PR08}.
The spin gauge field is introduced by diagonalizing the $sd$ interaction using a unitary transformation, $c(\bm{r},\tau)=U(\bm{r},\tau)a(\bm{r},\tau)$, where $U(\bm{r},\tau)$ is a $2\times2$ unitary matrix and $a$ is a new electron field operator.
A convenient choice of $U(\bm{r},\tau)$ is 
$U(\bm{r},\tau)=\bm{m}(\bm{r},\tau) \cdot \bm{\sigma}$ with $\bm{m}(\bm{r},\tau)=\left(\sin\frac{\theta}{2} \cos\phi , \sin\frac{\theta}{2} \sin\phi , \cos\frac{\theta}{2} \right)$, where $\theta$ and $\phi$ are the polar angles of $\bm{n}$. 
It is easy to confirm that $U^{\dagger}(\bm{n}\;\cdot\; \bm{\sigma})U$ = $\sigma_{z}$ is satisfied.
Because of this local unitary transformation, derivatives of the electron field become covariant derivatives  
$\partial_\mu c=U(\partial_\mu+ieA_{{\rm s},\mu})a$, where 
$A_{{\rm s},\mu}\equiv -\frac{i}{e}U^{-1}\partial_\mu U$ is the gauge field. 
Since $U$ is a $2\times2$ matrix, the gauge field $A_{{\rm s},\mu}$ is written using Pauli matrices as $A_{{\rm s},\mu}=\sum_{\alpha}A_{{\rm s},\mu}^\alpha \sigma_{\alpha}$ ($\mu=x,y,z,\tau$ is a suffix for space and time and $\alpha=x,y,z$ is for spin).
It is thus an SU(2) gauge field, which we call the spin gauge field.
The Lagrangian in the rotated space is thus given by
\begin{align}
\mathcal{L}&\equiv \mathcal{L}_{\text 0}+\mathcal{L}_{A},\\
\mathcal{L}_{\text 0}&\equiv \int {\rm{d}}^{3} r  \bar{a}  \left(\partial_\tau - \frac{1} {2m}  \bm{\nabla}^{2} - \mu - \Delta_{sd}\sigma_{z}\right) a, \\
\mathcal{L}_{A}&\equiv \int {\rm{d}}^{3} r  \left[ i e\bar{a} A_{{\rm s},\tau}a
+\sum_{i} \left(\sum_{\alpha} A_{{\rm s},i}^\alpha  j_{{\rm s},i}^\alpha +  A_{i}j_{i}\right) + \frac{e^2}{m}\sum_{i,\alpha}A_{i}A_{{\rm s},i}^\alpha \bar{a} \sigma^{\alpha}a\right.\nonumber\\
&\left.+ \frac{e^2}{2m}\sum_i\left(\sum_{\alpha} (A_{{\rm s},i}^\alpha) ^{2}+ (A_i)^{2}\right)\bar{a} a
\right], 
\end{align}
where 
$j_{{\rm s},i}^\alpha\equiv -ie\frac{1} {2m}\bar{a}\left(\overrightarrow{\nabla}_{i} - \overleftarrow{\nabla}_{i}\right)\sigma^\alpha a$
and $j_{i}\equiv -ie\frac{1} {2m}\bar{a}\left(\overrightarrow{\nabla}_{i} - \overleftarrow{\nabla}_{i}\right)a$ are the spin current and charge current, respectively.
The electron field $a$ is  strongly spin-polarized owing to the $sd$ exchange interaction (the last term of $\mathcal{L}_{\text 0}$).

We carry out the path integral with respect to the electron field and derive the effective Hamiltonian for the two gauge fields $A_{{\rm s},i}^\alpha$ and $A_{i}$ describing the spin and charge gauge fields, respectively.
The spin gauge field is written using the localized spin direction, $\theta$ and $\phi$, and thus the effective Hamiltonian can be regarded as that describing the interaction of the localized spin and the charge electromagnetic field.

Up to the second order with respect to the gauge fields, the effective Hamiltonian reads
\begin{align}
\mathcal{H}_{\rm eff} &= -\int_{0}^{\beta} {\rm{d}}\tau \int {\rm{d}}^3r \left[2ieA_{{\rm s},\tau}^{z}s_{\rm e} (\bm{r},\tau)+ \frac{2e^2}{m}\sum_{i}A_{i}A_{{\rm s},i}^{z}s_{\rm e}(\bm{r},\tau)\right.\nonumber\\
&\left.+ \frac{e^2}{2m}\sum_i\left(\sum_{\alpha} (A_{{\rm s},i}^\alpha) ^{2}+ (A_i)^{2}\right)n(\bm{r},\tau) \right]
+ \frac{1}{2}\int_{0}^{\beta} {\rm{d}}\tau\int_{0}^{\beta} {\rm{d}}\tau'\int {\rm{d}}^3r\int {\rm{d}}^3r'\sum_{ij} \nonumber\\
&\times \left[ \sum_{\alpha\beta} A_{{\rm s},i}^\alpha A_{{\rm s},j}^\beta \chi_{ij}^{\alpha\beta}(\bm{r},\bm{r}',\tau,\tau')
+ 2\sum_{\alpha} A_{{\rm s},i}^\alpha A_{j}\chi_{ij}^{\alpha}(\bm{r},\bm{r}',\tau,\tau')
+ A_{i} A_{j}\chi_{ij}(\bm{r},\bm{r}',\tau,\tau') \right],
\end{align}
where $s_{\rm e} (\bm{r},\tau)\equiv \frac{1}{2}\langle \bar{a}(\bm{r},\tau)\sigma^{z}a(\bm{r},\tau)\rangle$, $n(\bm{r},\tau)\equiv \langle \bar{a}(\bm{r},\tau)a(\bm{r},\tau)\rangle$
, $\langle\ \rangle$ denotes the thermal average, and
\begin{align}
\chi_{ij}^{\alpha\beta}(\bm{r},\bm{r}',\tau,\tau')
& \equiv \langle j_{{\rm s},i}^\alpha (\bm{r},\tau) j_{{\rm s},j}^\beta(\bm{r}',\tau') \rangle \nonumber\\
\chi_{ij}^{\alpha}(\bm{r},\bm{r}',\tau,\tau')
& \equiv \langle j_{{\rm s},i}^\alpha (\bm{r},\tau) j_{j}(\bm{r}',\tau') \rangle \nonumber\\
\chi_{ij}(\bm{r},\bm{r}',\tau,\tau')
& \equiv \langle j_{i} (\bm{r},\tau) j_{j}(\bm{r}',\tau') \rangle ,
\end{align}
are the current-current correlation functions.
The spin density $s_{\rm e}$ and the electron density $n$ are calculated as 
$
s_{\rm e}=\frac{1}{2V}\sum_{\kv}\sum_{\sigma=\pm}\sigma f(\epsilon_{\kv \sigma})$ and 
$n=\frac{1}{V}\sum_{\kv}\sum_{\sigma=\pm}f(\epsilon_{\kv \sigma})
$, respectively, 
where $f(\epsilon_{\kv \sigma})$=$(e^{\beta \epsilon_{\kv \sigma}}+1)^{-1}$ is the Fermi-Dirac distribution function, $\epsilon_{\kv\sigma}=\frac{k^2}{2m}-\mu-\sigma\Delta_{sd}$, and $\sigma$ = $\pm$ is the spin index. 
The Fourier components of the correlation functions are 
\begin{align}
\chi_{ij}^{\alpha\beta}(\qv,i\Omega_\ell)
& =  -\frac{e^2}{m^2\beta V}\sum_{n,\kv} k_i k_j {\rm tr}\left[\sigma_\alpha G_{\kv-\frac{\qv}{2},n} \sigma_\beta G_{\kv+\frac{\qv}{2},n+\ell}\right] 
\nonumber\\
\chi_{ij}^{\alpha}(\qv,i\Omega_\ell)
& =  -\frac{e^2}{m^2\beta V}\sum_{n,\kv} k_i k_j {\rm tr}\left[\sigma_\alpha G_{\kv-\frac{\qv}{2},n} G_{\kv+\frac{\qv}{2},n+\ell} \right]
\nonumber\\
\chi_{ij}(\qv,i\Omega_\ell)
& =  -\frac{e^2}{m^2\beta V}\sum_{n,\kv} k_i k_j {\rm tr}\left[G_{\kv-\frac{\qv}{2},n} G_{\kv+\frac{\qv}{2},n+\ell} \right] .
\end{align}
Here $G_{\kv,n}$ is defined as $G_{\kv,n}\equiv\left[i\omega_n-\epsilon_{\kv}+ \frac{i}{2\tau_e} {\rm sgn}(n)\right]^{-1}$,  where ${\rm sgn}(n)$ = $1$ and $-1$ for $n > 0$ and $n < 0$, respectively, 
$\epsilon_{\kv}=\frac{k^2}{2m}-\mu-\Delta_{sd}\sigma_z$ is the electron energy in the matrix representation, $\tau_e$ is the electron elastic scattering lifetime, 
and ${\rm tr}$ denotes the trace over spin space. 
The Fermionic thermal frequency is represented by $\omega_n$ $\equiv$ $\frac{(2n+1)\pi}{\beta}$, and $\Omega_\ell$ $\equiv$ $\frac{2\pi \ell}{\beta}$ is a bosonic thermal frequency. 

The correlation functions are calculated by rewriting the summation over the thermal frequency using the contour integral ($z$ $\equiv$ $i\omega_n$) as
\begin{align}
\chi_{ij}^{zz}(\qv,i\Omega_\ell)
& = \frac{e^2}{m^2V} \sum_{\kv} k_i k_j \sum_{\sigma=\pm}\int_C \frac{{\rm{d}}z}{2\pi i} f(z) g_{\kv-\frac{\qv}{2},\sigma}(z) g_{\kv+\frac{\qv}{2},\sigma}(z+i\Omega_\ell) 
\nonumber\\
\chi_{ij}^{+-}(\qv,i\Omega_\ell)
& =  \frac{e^2}{m^2V} \sum_{\kv} k_i k_j \sum_{\sigma=\pm}\int_C \frac{{\rm{d}}z}{2\pi i} f(z) g_{\kv-\frac{\qv}{2},\sigma}(z) g_{\kv+\frac{\qv}{2},-\sigma}(z+i\Omega_\ell) 
\nonumber\\
\chi_{ij}^{z}(\qv,i\Omega_\ell)
& =  \frac{e^2}{m^2V} \sum_{\kv} k_i k_j\sum_{\sigma=\pm}\int_C \frac{{\rm{d}}z}{2\pi i} \sigma f(z) g_{\kv-\frac{\qv}{2},\sigma}(z) g_{\kv+\frac{\qv}{2},\sigma}(z+i\Omega_\ell) 
\nonumber\\
\chi_{ij}(\qv,i\Omega_\ell)
& = \frac{e^2}{m^2V}\sum_{\kv} k_i k_j\sum_{\sigma=\pm}\int_C \frac{{\rm{d}}z}{2\pi i} f(z) g_{\kv-\frac{\qv}{2},\sigma}(z) g_{\kv+\frac{\qv}{2},\sigma}(z+i\Omega_\ell) ,
\end{align}
where $C$ is an anticlockwise contour surrounding the imaginary axis \cite{AGD75,Altland06} and $g_{\kv,\sigma}(z)$ is defined as $g_{\kv,\sigma}(z)\equiv \left[z-\epsilon_{\kv\sigma}+\frac{i}{2\tau_e}{\rm sgn}({\rm Im}z)\right]^{-1}$.

We expand the correlation functions with respect to the external wave vector $\qv$ and frequency $\Omega$ after the analytical continuation to $\Omega +i0$ $\equiv$ $i\Omega_\ell$ \cite{Altland06}. 
The result up to the second order in $\qv$ and $\Omega$ is
\begin{align}
\chi_{ij}^{\alpha\beta}(\qv,\Omega )
& = \frac{e^2}{m} \left\{ (\delta_{\alpha\beta} - \delta_{\alpha z}\delta_{\beta z})\delta_{ij} b  +\delta_{\alpha z}\delta_{\beta z}
\left[\delta_{ij} {n}\left(1+ i\Omega \tau_{\rm e} -(\Omega \tau_{\rm e})^2\right) 
 + c (q_iq_j - q^2\delta_{ij})\right]\right\}
\nonumber\\
\chi_{ij}^{z}(\qv,\Omega ) 
&=  \frac{e^2}{m} \left[\delta_{ij} 2s_{\rm e}\left(1+ i\Omega \tau_{\rm e} -(\Omega \tau_{\rm e})^2 \right) + 
 d (q_iq_j - q^2\delta_{ij})\right]
\nonumber\\
\chi_{ij}(\qv,\Omega )
& = \frac{e^2}{m} \left[\delta_{ij}n\left(1+ i\Omega \tau_{\rm e} -(\Omega \tau_{\rm e})^2\right) + 
 c (q_iq_j - q^2\delta_{ij})\right] ,
\end{align}
where $b\equiv \frac{1}{3mV\Delta_{sd}}\sum_{\kv}\sum_{\sigma=\pm}\sigma\kv^{2}f(\epsilon_{\kv\sigma}) $, $c \equiv \frac{1}{12m}\sum_{\sigma=\pm}\nu_\sigma$, 
$d \equiv \frac{1}{12m}\sum_{\sigma=\pm}\sigma\nu_\sigma$,
 $n=\sum_{\sigma=\pm}\frac{k_{{\rm F}\sigma}^2\nu_\sigma}{3m}$, 
$s_{\rm e}= \frac{1}{2}\sum_{\sigma=\pm}\sigma\frac{k_{{\rm F}\sigma}^2\nu_\sigma}{3m}$, 
and $k_{{\rm F}\sigma}\equiv \sqrt{k_{\rm F}^2+2m\sigma\Delta_{sd}}$ and $\nu_\sigma\equiv\frac{m^{\frac{3}{2}}}{\sqrt{2}\pi^2}\sqrt{\epsilon_{\rm F}+\sigma\Delta_{sd}}$ 
are the spin-dependent Fermi wave number and the density of states per unit volume, respectively. Vertex corrections are irrelevant, since they are proportional to $\bm{\nabla} \cdot \bm{E}$, which vanishes in metals. In this work, we do not consider surface effects played by induced surface charges such as the surface plasmon effect.
We thus obtain the effective Hamiltonian up to the order of ${\qv}^2$ and $\Omega^2$ as
\begin{align}
\mathcal{H}_{\text{eff}} &= - \Bigg\{ 
2ieA_{{\rm s},\tau}^z s_{\rm e} + \frac{1}{2m}\left(n-b\right)\sum_{i,\qv,\Omega} A_{{\rm s},i}^+(-\qv,-\Omega)A_{{\rm s},i}^-(\qv,\Omega) \nonumber\\
&-\frac{e^{2}}{m} 2s_{\rm e}\tau_{\rm e} \sum_{i,\qv,\Omega} i\Omega  A_{{\rm s},i}^z(-\qv,-\Omega)A_{i}(\qv,\Omega) \nonumber\\
&+ \frac{e^2}{2m}{\tau_{e}}^2\sum_{i,\qv,\Omega}\Omega^{2}\left[ n( A_{{\rm s},i}^z(-\qv,-\Omega)A_{{\rm s},i}^z(\qv,\Omega)+A_{i}(-\qv,-\Omega)A_{i}(\qv,\Omega) )
+ 4s_{\rm e} A_{{\rm s},i}^z(-\qv,-\Omega)A_{i}(\qv,\Omega) \right] \nonumber\\
&- \frac{e^2}{2m}\sum_{ij,\qv,\Omega}(q_iq_j - q^2\delta_{ij}) \nonumber\\
&
\times \left[ c ( A_{{\rm s},i}^z(-\qv,-\Omega)A_{{\rm s},j}^z(\qv,\Omega)+A_{i}(-\qv,-\Omega)A_{j}(\qv,\Omega) )
+ d  A_{{\rm s},i}^z(-\qv,-\Omega)A_{j}(\qv,\Omega) \right] \Bigg\},
\end{align}
where $A_{{\rm s},i}^{\pm}(\qv,\Omega)$ $\equiv$ $A_{{\rm s},i}^x(\qv,\Omega) \pm iA_{{\rm s},i}^y(\qv,\Omega)$.
The terms quadratic in the charge gauge field describe the electric permittivity and magnetic permeability of the media.
We used the fact that $\int dt A_{i}\dot{A}_{i}=0$ to drop the term proportional to $\Omega A_{i}(-\Omega)A_{i}(\Omega)$. 
The terms quadratic in the spin gauge field contribute to the renormalization of the exchange interaction and dissipation as shown in Ref. 35.
In fact, the contribution of the order of $\Omega^{0}$ reduces to
\begin{align}
\frac{1}{m}\left(n-b\right)\sum_{i} A_{{\rm s},i}^{+}A_{{\rm s},i}^{-}=J_{\text{eff}}(\bm{\nabla} \bm{n})^2,
\end{align}
where $J_{\text{eff}}\equiv \frac{1}{4m}\left(n-b\right)$. We treat this renormalization as a term in the original exchange interaction and do not consider it further. 
The term quadratic in the spin gauge field and linear in $\Omega$ represents a dissipation \cite{TF94_JPSJ}, but we neglect this effect since we are not interested in the spin dynamics where dissipation plays an important role.
Instead, we are interested in the coupling between the two gauge fields. 
The effective Hamiltonian describing the coupling reads 
\begin{align}
\mathcal{H}_{\rm int} &= \frac{e^2}{m} \int {\rm{d}}^3r \left(
 2s_{\rm e} \tau_{\rm e} \bm{E}\cdot\bm{A}_{{\rm s}}^z+ 2s_{\rm e} \tau_{\rm e}^2 \bm{E}\cdot\bm{E}_{{\rm s}}
  + \frac{d}{2} \bm{B}\cdot\bm{B}_{\rm s}       \right),
\label{Hcoupling}
\end{align}
where $\bm{E}\equiv -\dot{\bm{A}}$ and $\bm{B}\equiv \nabla\times \bm{A}$ are the electric and magnetic fields, respectively, and $\bm{E}_{\rm s}\equiv -\dot{\bm{A}}_{\rm s}^z$ and  $\bm{B}_{\rm s}\equiv \nabla\times\bm{A}_{\rm s}^z$ are the effective spin electric and magnetic fields, respectively.

\section{Discussion}

Let us discuss the effect of the coupling terms, Eq. (\ref{Hcoupling}).
The first term indicates that $\bm{A}_{\rm s}^z$ is induced when an electric field is applied.
In fact, this term is the term describing the spin-transfer torque, 
as seen by denoting  
$2s_{\rm e}  \frac{e^2}{m} \tau_{\rm e} \bm{E}=P\bm{j}$, where $P\equiv \frac{n_\uparrow-n_\downarrow}{n}$, $\bm{j}=\sigma_{\rm B}\bm{E}$, and $\sigma_{\rm B}\equiv  \frac{e^2}{m}n \tau_{\rm e}$ is the Boltzmann conductivity.
As pointed out in Ref. 6, the effective Hamiltonian method that we used thus easily reproduces the spin-transfer effect, which is usually discussed in the context of the conservation law of angular momentum.
Although the spin gauge field $\bm{A}_{\rm s}^z$ is related to the spin electric field as 
$\bm{E}_{\rm s}=-\dot{\bm{A}}_{\rm s}^z$, the generation of $\bm{A}_{\rm s}^z$ does not always imply the generation of a spin electric field. 
In fact, a direct consequence of the spin-transfer torque is to drive magnetization textures \cite{TKS_PR08}. 
Only when the induced magnetization dynamics creates a non-coplanarity, the spin electric field is induced. 
The emergence of the spin electric field thus depends in an essential way on the dynamics of the magnetization.
\subsection{Spin electric field induced by domain wall motion}
Let us consider as an example a domain wall.
A domain wall favors a non-coplanar motion since its center of mass coordinate $X$ and the angle of the wall plane $\phi$ are canonical conjugates to each other \cite{TKS_PR08}. 
When the spin-transfer torque due to an electric field is applied, the wall plane starts to tilt and its angle drives the motion of the wall \cite{TK04}.
The spin electric field generated by this wall motion is calculated as follows.
We consider a case of uniaxial anisotropy and neglect the nonadiabaticity which is represented by $\beta$ in Ref. 5. 
A planar domain wall with the magnetization changing along the $x$-direction at position $x=X(t)$ is described by $\cos\theta=\tanh\frac{x-X}{\lambda}$ and 
$\sin\theta=\frac{1}{\cosh\frac{x-X}{\lambda}}$ with a constant $\phi$.
The equations of motion for $X$ and $\phi$ are \cite{TTKSNF06}
\begin{align}
\dot{X}-\alpha\lambda \dot{\phi} 
&=
\frac{a_0^3}{2eS}Pj \nonumber \\
\dot{\phi}+\alpha\frac{\dot{X}}{\lambda}
&= 0, \label{DWeq}
\end{align}
where $\lambda$ is the width of the wall, $\alpha$ is the Gilbert damping parameter, $a_0$ is the lattice constant,
and $S$ is the magnitude of the localized spin. 
The solution for Eq. (\ref{DWeq}) is
$\dot{X}=\frac{1}{1+\alpha^2}\frac{a_0^3}{2eS}Pj$ and 
$\dot{\phi}=-\frac{\alpha}{1+\alpha^2}\frac{a_0^3}{2eS\lambda}Pj$.
The spin electric field, Eq. (\ref{EBtopological}), for a moving domain wall is 
calculated using 
$\nabla_x{\nv}=-\frac{\sin\theta}{\lambda}\ev_\theta$ and 
$\dot{\nv}=\sin\theta\left(\dot{\phi}\ev_{\phi}+\frac{\dot{X}}{\lambda}\ev_\theta\right)$,
where $\ev_\phi=(-\sin\phi,\cos\phi,0)$ and $\ev_\theta=(\cos\theta\cos\phi,\cos\theta\sin\phi,-\sin\theta)$.
The spin electric field then arises along the $x$-direction and the magnitude is  
\begin{align}
E_{\rm{s}}&=
\frac{\hbar}{2e} \frac{1}{2\lambda}  \dot{\phi}=  \frac{\hbar}{2e^2}\frac{\alpha}{1+\alpha^2}\frac{a_0^3P}{4S\lambda^2}\sigma_{\rm B}E.
\label{EsDW}
\end{align}
Let us estimate the magnitude choosing $\alpha \sim 10^{-2}$, $P \sim 0.8$ \cite{Ueda12}, $\sigma_{\rm B} \sim 10^{8}$ $\Omega^{-1}\text{m}^{-1}$, and $S = 1$. For a field of  $E\sim 10^{4}$ $\text{V}/\text{m}$ corresponding to $j=10^{12}$ A/m$^2$, we have $|\dot{X}|$ $\sim$ $4$ $\text{m}/\text{s}$ and $|\dot{\phi}|\sim4\times10^{6}$ $\text{s}^{-1}$.
For  $\lambda$ $\sim$ $10^{-8}$ $\text{m}$ \cite{Fukami08}, we thus obtain $|E_{\rm s}|$ $\sim$ $0.1$ $\text{V}/\text{m}$. In experiments, what is measured is the voltage due to $E_{\rm s}$. Since $E_{\rm s}$ is localized at the wall, the voltage generated by a single domain wall is $E_{\rm s}\lambda \sim 1 \;\text{nV}$. This value is not large, but is detectable. For instance, in the case of a domain wall driven by an external magnetic field, a voltage of $400\;\text{nV}$ was observed for a wall speed of $150\;\text{m}/\text{s}$ \cite{Yang09}.
The conversion efficiency from the electric field to the spin electric field is given by
\begin{equation}
\mu\equiv \frac{E_{\rm s}}{E}=\frac{\hbar}{2e^2}\frac{\alpha}{1+\alpha^2}\frac{a_0^3P}{4S\lambda^2}\sigma_{\rm B}
\sim \frac{P\alpha}{4(k_{\rm F}\lambda)^2} \left(\frac{\epsilon_{\rm F}\tau_{\rm e}}{\hbar}\right),
\end{equation}
where we approximated $s_{\rm e}a_0^3 \sim P/2$. A typical value of $\mu$ is $\mu \sim 10^{-4}$ for $\alpha \sim 10^{-2}$, $k_{\rm F}\lambda \sim 100$, and $(\epsilon_{\rm F}\tau_{\rm e})/\hbar \sim 100$. The conversion efficiency is larger in a thin wall, such as perpendicular anisotropy magnets and weak ferromagnets. 

In contrast to the spin-transfer term, the second term in Eq. (\ref{Hcoupling}) describes a direct coupling between the electric field and the spin electric field.
The strength of the induced spin electric field is determined by solving spin dynamics as we did above, because  $E_{\rm s}$ is determined using the equation of motion for the spin (the Laudau-Lifshitz equation) and not for the Lagrangian like $(\bm{E}_{\rm s})^2-(\bm{B}_{\rm s})^2$ as in the charge electromagnetism.
The effect is generally small, since the $\bm{E}\cdot\bm{E}_{\rm s}$ coupling term is smaller than the spin-transfer term by a factor of $\Omega\tau_{\rm e}$, which is small in the GHz frequency where magnetization structures can respond [$\tau_{\rm e}\gtrsim 10^{-13}$ $\text{s}$ for a metal with $(\epsilon_{\rm F}\tau_{\rm e})/\hbar = 100$].
The term may be important in a very clean metal in the $\text{THz}$ range.

 \subsection{Magnetic coupling}
 The third term of Eq. (\ref{Hcoupling}) is due to the external magnetic field.
 It indicates that the spin magnetic field is induced when a uniform external magnetic field is applied.
This effect indicates a novel intrinsic mechanism involving frustration, since a finite $\bm{B}_{\rm s}$ denotes a non-coplanar spin structure, while a uniform magnetic field favors the uniform magnetization along the magnetic field.
When the magnetization structure has a finite non-coplanarity at the scale of $\lambda$, the magnitude of the induced spin magnetic field is  ${B}_{\rm s}\sim \frac{\hbar}{e \lambda^2}$.
The energy gain per site due to the non-coplanarity is then 
$\left(\frac{e\hbar}{m}\right)^{2}\frac{{B}_{\rm s}}{\epsilon_{\rm F}}B\simeq \frac{e\hbar}{m}B \frac{2}{(k_{\rm F} \lambda)^2}$, assuming that $d \sim O(\frac{1}{12m\epsilon_{\rm F}})$. [Note that $d=\frac{1}{12m}(\nu_{+}-\nu_{-})$ can be positive or negative.]
In ferromagnetic systems, a non-coplanar structure costs the exchange energy of 
$\frac{J}{2}(\nabla \nv)^2\simeq \frac{J}{a_0^2}  \frac{1}{(k_{\rm F} \lambda)^2}$, where $J$ is the exchange energy of the localized spin, 
in the absence of other frustration.
The total energy cost due to the generation of $\bm{B}_{\rm s}$ is thus
\begin{align}
  \Delta E\simeq   \frac{1}{(k_{\rm F} \lambda)^2} \left( \frac{J}{a_0^2} - \frac{e\hbar}{m}B \right).
\end{align}
In common $3d$ ferromagnetic metals, the exchange energy $J/a_0^2$ is on the order of 1 $\text{eV}$, while the applied magnetic field of 1 $\text{T}$ corresponds to the energy of $\frac{e\hbar}{m}B=10^{-4}$ $\text{eV}$. 
The spin magnetic field $\bm{B}_{\rm s}$ is therefore not induced by simply applying a uniform magnetic field.
The situation may be different in very weak ferromagnets like in molecular conducting ferromagnets.
For a system with a ferromagnetic critical temperature of 5 $\text{K}$  \cite{Coronado04}, a magnetic field of 5 $\text{T}$ may be sufficient to induce a finite $\bm{B}_{\rm s}$ if the Fermi energy is on the order of 1 $\text{eV}$. Our present study assuming a strong $sd$ exchange interaction does not directly apply to weak ferromagnets, and different approaches like in Ref. 21 are needed to study the interplay between $\bm{B}_{\rm s}$ and $\bm{B}$.
Molecular conducting ferromagnets would be unique systems in the context of an emergent spin electromagnetic field.

\subsection{Spin electric field induced by spin wave}
The spin-transfer term $\mathcal{H}_{\rm st}$ induces a spin wave excitation when the electric field is applied if the induced current exceeds a threshold value \cite{STK05}.
If the spin wave is monochromatic, a spin electric field is not induced, since a monochromatic plane wave, such as $n_x\pm i n_y=\varphi e^{\pm i(kx-\Omega t)}$, does not have a non-coplanarity because $\dot{\bm{n}}$ and $\nabla_x\bm{n}$ are parallel to each other.
It is possible to excite a spin electric field if we use two spin waves having different wave vectors or frequencies.
Let us consider the case described by 
\begin{align}
s_{\pm}=\frac{1}{2}\sum_{j=1,2} \varphi_j e^{\pm i(k_j x-\Omega_j t)},                                    
\end{align}
where $s_{\pm}\equiv \frac{1}{2}(s_x\pm i s_y)$ is a small spin fluctuation, $\varphi_j$ are the amplitudes of spin waves, and $k_j$ and $\Omega_j$ ($j=1,2$) are the wave vector and angular frequency of the two spin-wave excitations, respectively. 
The spin electric field associated with the spin waves then reads 
\begin{align}
E_{{\rm s},x}=-\frac{\hbar}{2e}\varphi_1 \varphi_2 (\Omega_1k_2-\Omega_2k_1)\sin[(k_1-k_2)x-(\Omega_1-\Omega_2)t].
\end{align}
Namely, a spin electric field having a wave vector $(k_1-k_2)$ and a frequency $(\Omega_1-\Omega_2)$ is induced  by the two spin-wave excitations. Two spin waves having the same frequency of about 7 $\text{GHz}$ have been generated recently using a magnetic field which is induced by applying a current through antennas \cite{Sato13}. 
The method would be applicable to create a spin electric field.
For spin waves with $k=0.4$ $\mu$m$^{-1}$ and an angular frequency of $2\pi \times7$ $\text{GHz}$ \cite{Sato13}, $k\Omega\sim 1.7\times 10^{15}$ 1/(ms), and the spin electric field is expected to be on the order of $\frac{\hbar}{e}k\Omega\varphi^2\simeq 1.7 \times  \varphi^2$ V/m for the amplitude of spin waves $\varphi$. The spin electric field is expected to be induced generally by the spin wave excitation owing to a nonlinear effect ($E_{{\rm s},x}$ nonlinearly depend on the spin wave amplitude), if the wave is non-monochromatic.
The present method of generating a spin electric field applies to a uniform ferromagnet and would have better possibilities of applications than the conventional methods using magnetic structures such as domain walls. 

\section{Summary}
We have derived the effective Hamiltonian describing the coupling between the emergent spin electromagnetic field and the charge electromagnetic field. The dominant term turns out to be the one corresponding to the spin-transfer torque. The coupling between the magnetic components suggests an interesting possibility of inducing frustration by applying a uniform external magnetic field on weak ferromagnets.
We have proposed a generation mechanism of a spin electric field using a nonlinear effect of non-monochromatic spin-wave excitations.
This mechanism is applicable to the case of a uniform magnetization, and it would have a great advantage in applications over common setups using non-coplanar structures. 
Our theoretical considerations call for an experimental verification of the effect.

\section*{Acknowledgments}
The authors thank N. Nakabayashi, H. Kohno, H. Saarikoski, and H. Seo for valuable comments and discussions. 
This work was supported by a Grant-in-Aid for Scientific Research (C) (Grant No. 25400344) and (A) (Grant No. 24244053) from the Japan Society for the Promotion of Science and UK-Japanese Collaboration on Current-Driven Domain Wall Dynamics from JST.

\end{document}